\documentclass[secnumarabic,twocolumn,nofootinbib,tightenlines,showpacs,aps,superscriptaddress]{revtex4}
\usepackage{amsfonts}
\usepackage{amsmath}
\usepackage{amssymb}
\usepackage{graphicx}%
\usepackage{bm}%
\usepackage{setspace}
\usepackage{longtable}\makeatletter
\let\@currsize\normalsize
\makeatother
\usepackage{setspace}
\begin{document}

\def\Fig#1{Figure \ref{#1}}
\def\Eq#1{Eq.~(\ref{#1})}
\def\Table#1{Table~\ref{#1}}

\title{Overlimiting current in a microchannel}

\author{E. Victoria Dydek}
\affiliation{Department of Chemical Engineering, Massachusetts Institute of Technology, Cambridge, MA 02139 USA}
\author{Boris Zaltzman}
\affiliation{Department of Chemical Engineering, Massachusetts Institute of Technology, Cambridge, MA 02139 USA}
\affiliation{Blaustein Institutes for Desert Research, Ben-Gurion University of the Negev, Sede Boqer Campus, 84990, Israel}
\author{Isaak Rubinstein}
\affiliation{Blaustein Institutes for Desert Research, Ben-Gurion University of the Negev, Sede Boqer Campus, 84990, Israel}
\author{D. S. Deng}
\affiliation{Department of Chemical Engineering, Massachusetts Institute of Technology, Cambridge, MA 02139 USA}
\author{Ali Mani}
\affiliation{Department of Chemical Engineering, Massachusetts Institute of Technology, Cambridge, MA 02139 USA}
\author{Martin Z. Bazant }
\affiliation{Department of Chemical Engineering, Massachusetts Institute of Technology, Cambridge, MA 02139 USA}
\affiliation{Department of Mathematics, Massachusetts Institute of Technology, Cambridge, MA 02139 USA}
\date{\today}

\begin{abstract}{ We revisit the classical problem of diffusion-limited ion transport to a membrane (or electrode) by considering the effects of charged side walls. Using simple mathematical models and numerical simulations, we identify three basic  mechanisms for over-limiting current in a microchannel: (i) {\it surface conduction} carried by excess counterions, which dominates for very thin channels, (ii) convection by {\it electro-osmotic flow} on the side walls, which dominates for thicker channels and transitions to (iii) {\it electro-osmotic instability} on the membrane end in very thick channels. These intriguing electrokinetic phenomena may find applications in biological separations, water desalination, and electrochemical energy storage. }
\end{abstract}

\maketitle

\emph{Introduction. ---} All electrochemical cells involve ion-selective surfaces, such as electrodes or  membranes, that pass current carried by certain active ionic species~\cite{gen}.
In an unsupported bulk electrolyte, the  rejection of the opposite species (to maintain neutrality) leads to salt depletion and ``concentration polarization" (CP). Theoretically, at the diffusion-limited current, the solution conductivity vanishes at the surface,  but in practice, ``overlimiting current" (OLC) is often observed and associated with an extended depletion zone (Fig. 1a).

Possible mechanisms for OLC have been debated for decades. In the context of water electrodialysis~\cite{nik}, it is now understood that OLC can be due to chemical effects (water splitting or exaltation), which create additional ions, or a physical effect, electro-osmotic instability (EOI)~\cite{1}, which enhances ion transport by convection~\cite{2} (Fig. 1d). The flow is driven by extended space charge (ESC) in the non-equilibrium double layer on the selective surface~\cite{esc}.
Similar phenomena can also occur in dendritic electrodeposition~\cite{ed}, electrochromatography~\cite{tallarek} and capacitive desalination at large voltages~\cite{cd}.

Some experiments have shown that CP is also affected by {\it geometrical confinement}. In thin-gap ($100\mu$m) electrodeposition cells, gravitational and electro-osmotic convection can affect the spreading depletion layer~\cite{ed}. In thinner microfluidic devices, steady depletion layers in transverse flows can form surprisingly sharp interfaces (``demixing"), whether triggered by capacitive desalination~\cite{leinweber2006} or electrodialysis~\cite{han}. Additionally, convection can be suppressed by decreasing the channel height, thereby reducing (but not eliminating) OLC~\cite{kim2007,yossifon2010}. In very thin ($1 \mu$m) channels, it was recently shown that CP interfaces can propagate at constant current, analogous to shock waves in gases~\cite{3}. These ``desalination shocks"  result from geometry-dependent surface conduction (Fig. 1b) ~\cite{mb}, but the role of convection and the connection with OLC have not been considered.

\begin{figure}
\includegraphics[width=\columnwidth,keepaspectratio=true]{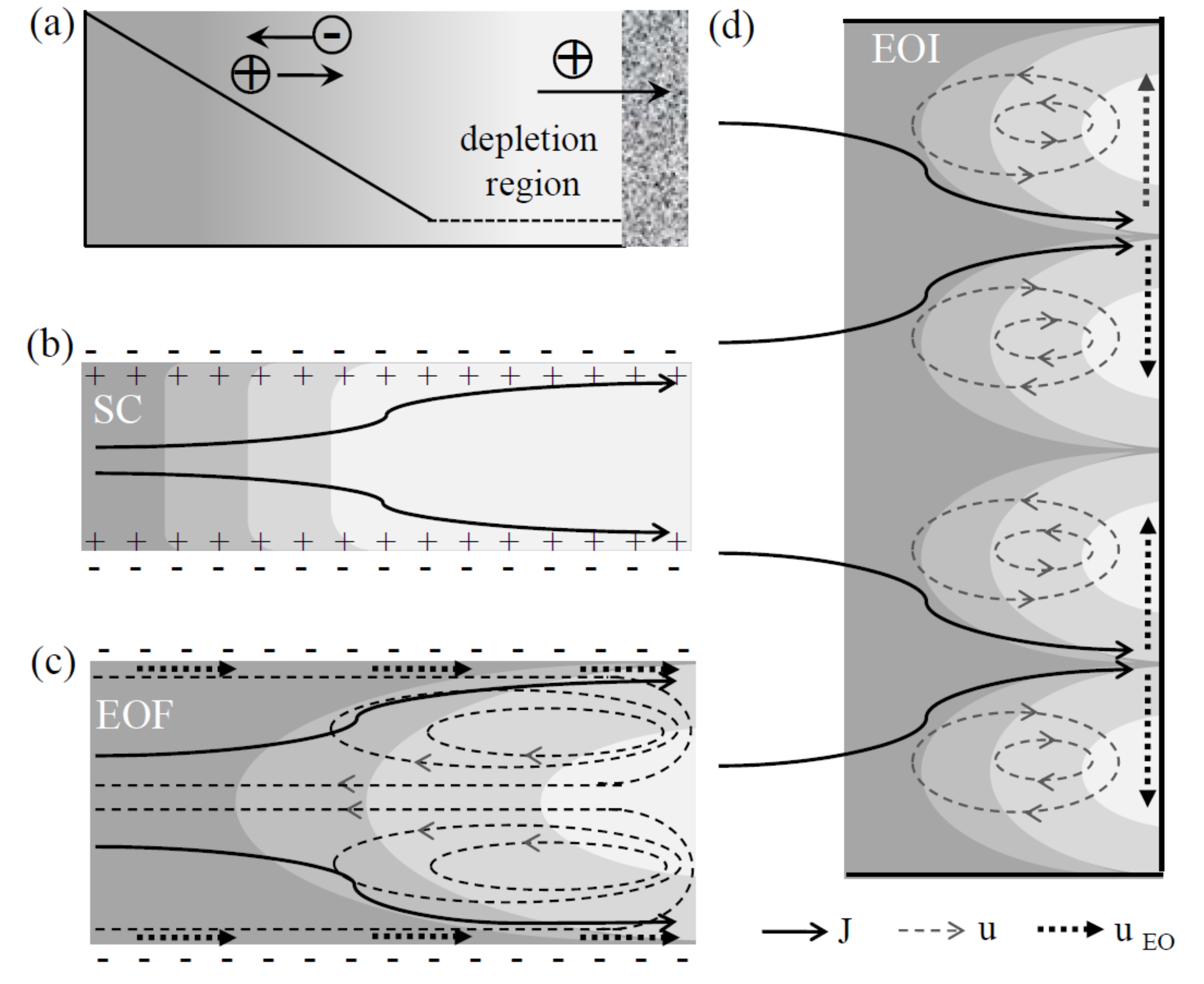}
\caption{ Physical picture of over-limiting current (OLC) in a microchannel from a reservoir (left) to a cation-exchange membrane (right). The volume-averaged conductivity profile (a) exhibits a classical linear diffusion layer (continuous line) and a constant depleted region (dashed line), where the current is carried primarily by (b) SC, (c) EOF or (d) EOI with increasing channel thickness. }
\end{figure}

In this Letter, we develop a unified theory of steady CP in a microchannel (neglecting chemical effects) and identify physical mechanisms for OLC related to {\it charged side walls}. We show that surface conduction (SC) dominates in very narrow channels (Fig. 1b), while electro-osmotic flow (EOF) on the side walls dominates in wider channels (Fig 1c) and transitions to EOI on the selective surface in very wide channels (Fig. 1d). We are not aware of any prior mention of the SC mechanism for OLC,  whereas our theory of the EOF mechanism differs from the recently proposed 1D Taylor-Aris dispersion model (strictly valid for aspect ratios $>10^3$ and zeta potentials $> 1$ V)~\cite{4}, by taking into account the 2D velocity field at the channel end.

Our analysis is based on the following model problem. A symmetric, binary $z:z$ electrolyte fills a parallel-plate (2D) microchannel of length $L$ and thickness $H$, which is open to a well-mixed reservoir of concentration $c_0$ at $x=0$. A voltage $V$ is applied across the microchannel to drive a current $I$ through an ideal cation-selective solid surface (the ``membrane") at $x=L$, which could represent a cation-exchange membrane, a negatively charged nanochannel, or a metal electrode undergoing electroplating. The side walls have a charge density $\sigma_s<0$, which we take to be negative, as in most materials used in microfabrication (glass, silicones, silica, etc.). We shall see that the surface promotes OLC only if its charge has the same sign as the inactive species (anions).

\emph{ Surface Conduction. (SC)} ---  As the channel thickness is decreased, convection eventually becomes negligible compared to diffusion (small P{\' e}clet number), so we begin by considering only ``surface conduction'' associated with the excess counterions (active cations) that screen the wall charge~\cite{sc,4}, neglecting streaming current (which we analyzed separately~\cite{epaps}). For long, narrow channels ($H \ll L$), and thin double layers ($\lambda_D \ll H$, where $\lambda_D$ is the Debye screening length), the Nernst-Planck equations can be homogenized (area-averaged) as follows~\cite{helfferich,5,mb},
\begin{align}
&\frac{dc_+}{dx}+c_+\frac{d\tilde{\phi}}{dx}=-\frac{j}{zeD},\ \frac{dc_-}{dx}-c_-\frac{d\tilde{\phi}}{dx}=0,\label{1}\\
& c=c_-=c_++\frac{\rho_s}{e},\label{2}
\end{align}
where $c_+$ and $c_-$ are the mean concentrations of cations and anions, respectively, $D$ is the ionic diffusivity (for simplicity, the same for both species), $\tilde{\phi}$ is dimensionless potential scaled to the thermal voltage, $kT/ze$.  Equation (\ref{1}) relates the cation flux to the current density $j$ in steady state for a perfect cation-selective surface. Equation (\ref{2}) is a mean electroneutrality condition including both ionic and fixed surface charges, where $\rho_s=2\sigma_s/H$ is the negative volume density of fixed charge. 

The homogenized 1D model can be solved analytically:
\begin{eqnarray}
 \tilde{\phi}=\ln(\tilde{c}),\ \ \  \tilde{c}-\tilde{\rho}_s\ln(\tilde{c})=1-\tilde{j}x,\label{3}\\
\tilde{j}=1-e^{-\tilde{V}}-\tilde{\rho}_s\tilde{V}, \ \ \ \mbox{(SC)} \label{4}\
\end{eqnarray}
where $\tilde{\rho}_s=\frac{\rho_s}{2zec_0}$ is dimensionless fixed-charge density, $\tilde{j}=\frac{j L}{2zeDc_0}$ is the dimensionless current density scaled by the limiting current, corresponding to the case with neutral side walls; $\tilde{c}=c/c_0$, $\tilde{x}=x/L$, and voltage $\tilde{V}=zeV/kT=-\tilde{\phi}(1)$ is the potential drop across the electrolyte, relative to $\tilde{\phi}(0)=0$ at the reservoir. The current-voltage relationship, Eq. (4), is shown in Fig. 2(a) demonstrating nearly constant conductance in the OLC regime. For small surface charge and/or wide channels, $0<-\tilde{\rho}_s\ll 1$, the anion concentration profile (\ref{3}) is linear in the quasi-neutral bulk region, $\tilde{c}\sim 1-\tilde{j}x$,  and decays exponentially, $\tilde{c} \sim e^{(\tilde{j}x-1)/\tilde{\rho}_s}$, in the depleted zone, $\tilde{j}^{-1} < x < 1$, that forms above the limiting current, as shown in Fig. 2(b). In the opposite ``membrane limit", $-\tilde{\rho}_s \gg 1$, the concentration is only weakly perturbed by the current, $\tilde{c} \sim 1 + (\tilde{j}x-1)/\tilde{\rho}_s$.


\begin{figure}
\includegraphics[width=1.65in]{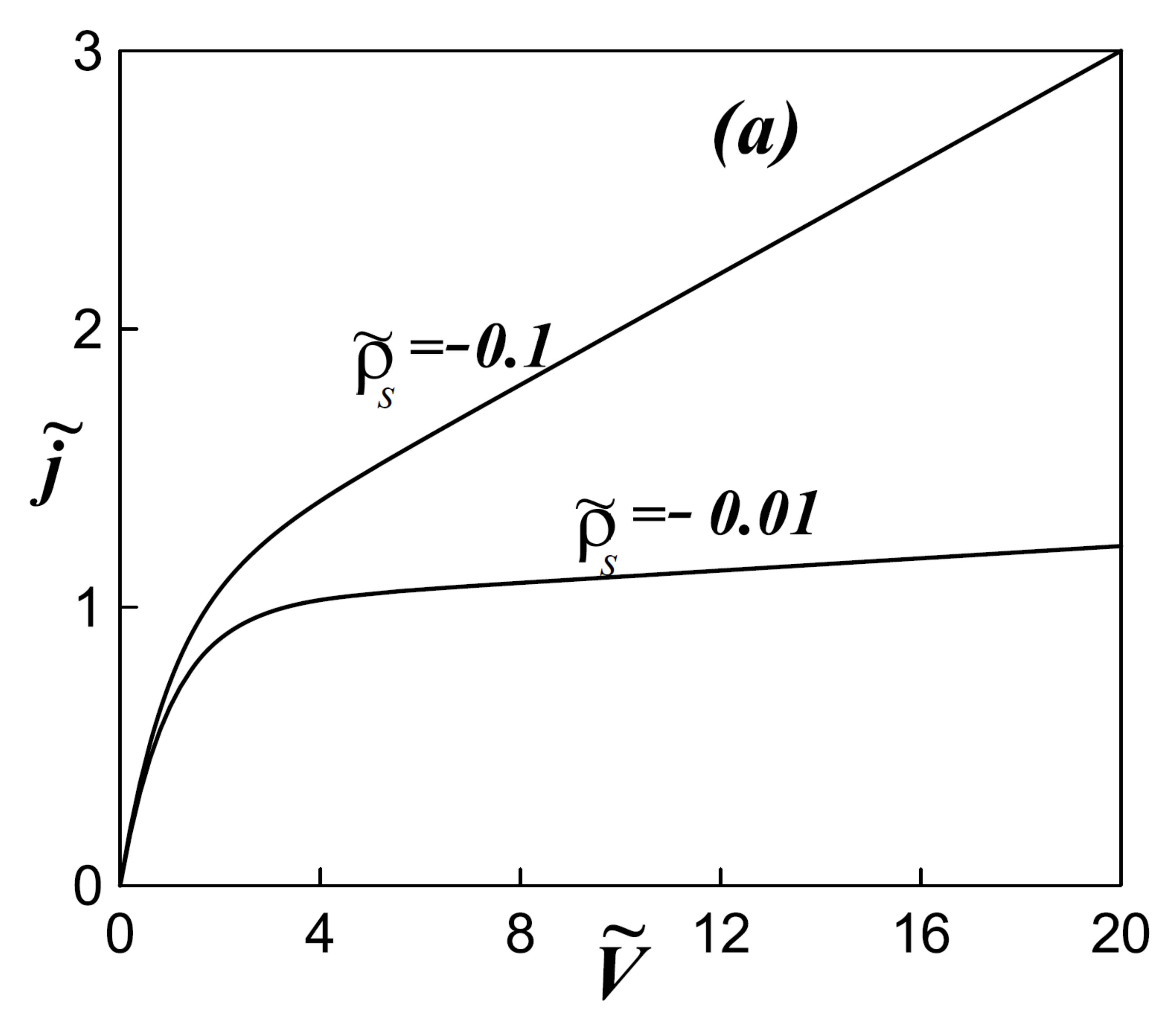} \hspace{-.05in}
 \includegraphics[width=1.65in]{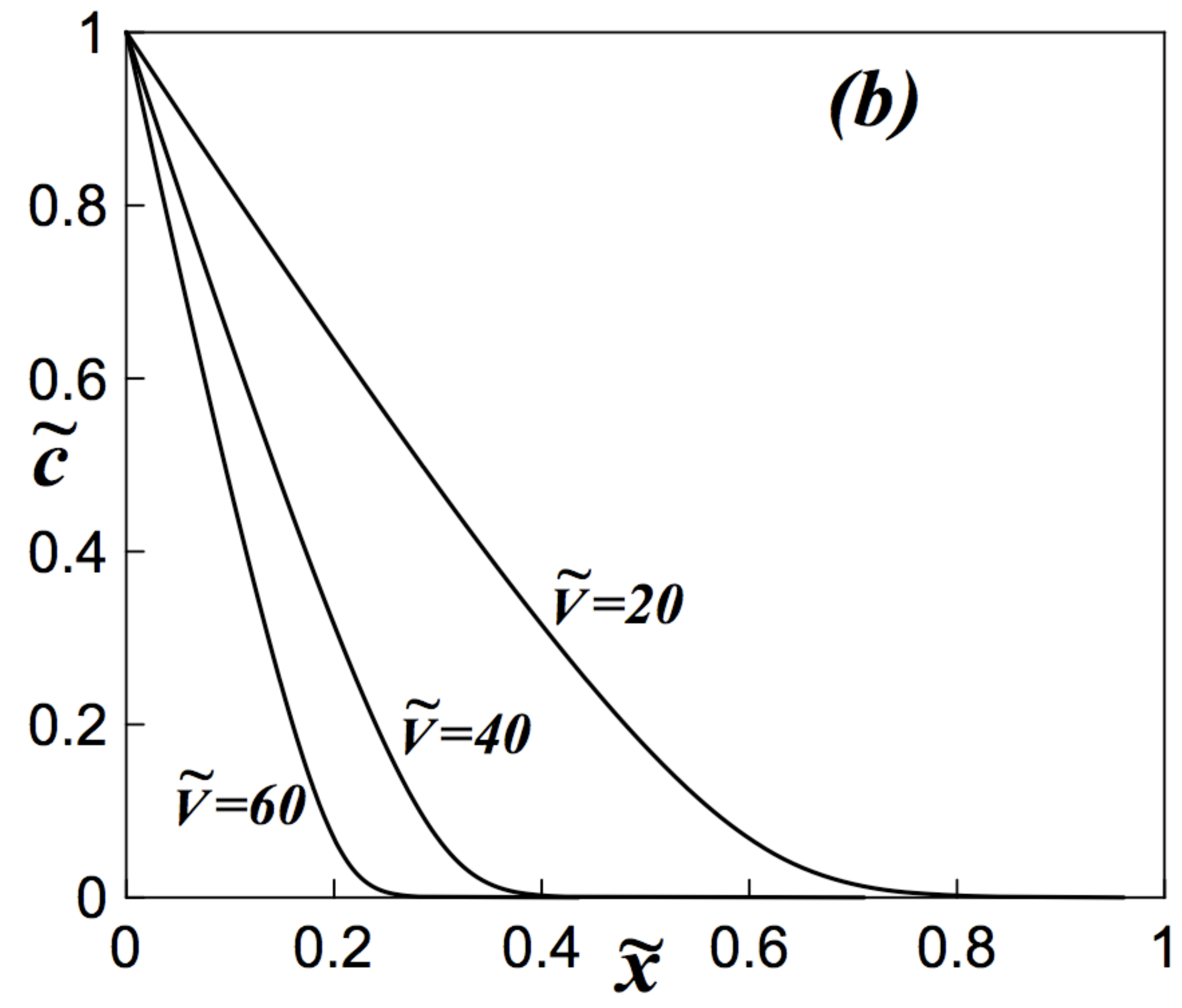} \\
 \vspace{0.1in}
(c)\includegraphics[width=2.5in]{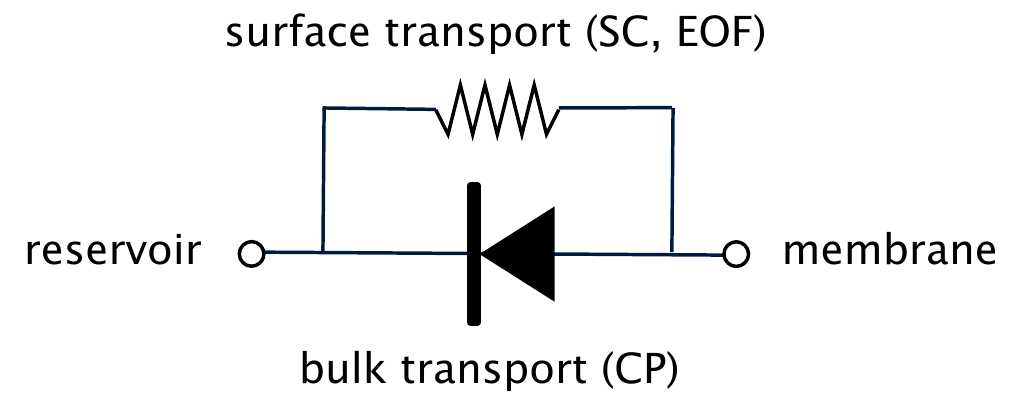} \\
\caption{ Overlimiting current in the surface conduction regime: (a) Dimensionless current-voltage relation, Eq. (\ref{4}), for $\tilde{\rho}_s=-0.01,-0.1$ and (b) mean concentration profile of inert anions, Eq. (\ref{3}), for $\tilde{\rho}_s=-0.1$. (c) The equivalent circuit: an ideal diode for bulk CP in parallel with a constant shunt resistance for SC (or EOF, as in Fig. 3).  }
\end{figure}

As shown in Fig. 2(c), the current-voltage relation (\ref{4}) can be interpreted as an ideal diode, $\tilde{j}_b=1-e^{-\tilde{V}}$, for bulk CP in parallel with a shunt resistance, $\tilde{j}_s=-\tilde{\rho}_s\tilde{V}$, for SC. The electric field, which is nearly uniform across the microchannel, acts on two different types of ions: (i) the inert anions plus an equal number of cations (``bulk conductivity"), and (ii) the remaining cations that screen the surface charge (``surface conductivity").  The response is equivalent to that of an ideal Schottky diode (metal/semiconductor junction)~ \cite{sze} with the surface charge playing the role of a n-type dopant in the semiconductor, which provides a residual conductivity under reverse bias. As the microchannel thickness is increased, convection due to EOF eventually dominates SC, and the semiconductor analogy breaks down.
\begin{figure*}
\includegraphics[width=\linewidth,keepaspectratio=true]{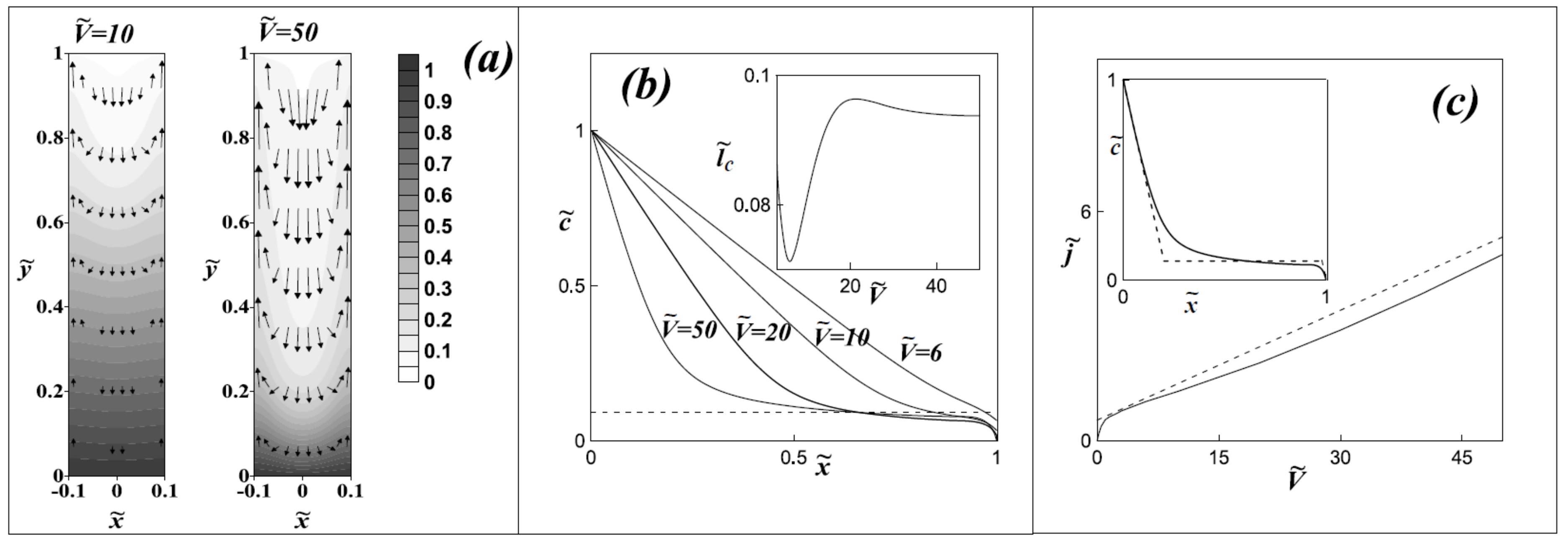}
\caption{ 2D simulations of the EOF mechanism for OLC ($\tilde{\lambda}_0=0.01,\ \tilde{H}=0.2, \tilde{\sigma}_s=-0.01$): (a) Bulk salt concentration $\tilde{c}$ (contours) and velocity vectors $\mathbf{u}$ (arrows); (b) Area-averaged concentration profiles (continuous), showing a nearly constant depleted-zone concentration $\tilde{c}_d\approx\tilde{l}_c$. Inset: vortex position $\tilde{l}_c$ vs. voltage $\tilde V$; (c) Current-voltage relation from simulations (continuous) and the 1D model (\ref{7}) (dashed). Inset: Area-averaged concentration profile in the 2D simulations (continuous) and in 1D model (dashed) for $\tilde{V}=50$ .}
\end{figure*}

\emph{Electro-osmotic Flow (EOF).} --- For our ``dead-end channel" with an impermeable membrane,   pressure-driven back flow opposes EOF and results in two counter-rotating vortices. Concentrated solution flows to the membrane along the side walls, and depleted solution returns in the center (Fig. 1(c)).

As a first approximation, we assume thin double layers and a neutral bulk solution ($c=c_-=c_+$) described by the steady 2D Nernst-Planck-Stokes equations,
\begin{eqnarray}
\mathbf{u}\cdot\mathbf{\nabla} c=-\mathbf{\nabla}\cdot\mathbf{F}_\pm, \ \ \ \
\mathbf{F}_\pm=-D\left(\nabla c \pm c \mathbf{\nabla}\tilde{\phi}\right) \label{5} \\
\nabla p=\eta\mathbf{\nabla}^2\mathbf{u}, \ \ \ \ \mathbf{\nabla}\cdot\mathbf{u}=0
\end{eqnarray}
where $\mathbf{u}$ is the fluid velocity, $\eta$ the viscosity, and $p$ the pressure with boundary conditions of no slip, $\mathbf{u}=\mathbf{0}$, and vanishing normal co-ion flux and specify uniform electrochemical potential of counter-ions at the membrane corresponding to potentio-static operating conditions; uniform concentration $c(0,y)=c_0$ and zero pressure at the reservoir; and zero normal flux (neglecting SC) and EOF slip at $y=\pm H/2$ given by the Helmholtz-Smoluchowski formula (neglecting diffusio-osmosis~\cite{2}),
\begin{equation}
\mathbf{u}_x= \frac{\varepsilon\zeta}{\eta} \frac{\partial \phi}{\partial x}, \ \ \zeta=\frac{\sigma_s\lambda_D}{\varepsilon},
\ \ \lambda_D(c) = \sqrt{ \frac{\varepsilon k_B T}{2 (ze)^2 c} } \label{6}
\end{equation}
where $\zeta(c)$ is the zeta potential in the Debye-H\"uckel approximation and $\varepsilon$, the permittivity. The main output parameter in model  (\ref{5})-(\ref{6}), the dimensionless current, $\tilde{j}$, depends on the dimensionless channel thickness, $\tilde H = \frac{H}{L}$, reservoir Debye length, $\tilde{\lambda}_0=\frac{\lambda_D(c_0)}{L}$, surface charge, $\tilde{\sigma}_s=\tilde{\rho}_s\tilde{H}$,  and voltage, $\tilde V=\frac{zeV}{kT}$.


Numerical solutions~\cite{epaps} of (\ref{5})-(\ref{6})  are shown in Fig. 3. An interesting observation about CP during OLC follows from averaging the 2D concentration profile (Fig. 3a) over the channel cross section (Fig. 3b). The intense vortical circulation near the dead end forms an extended depletion zone with a nearly constant area-averaged concentration,  $c_d=\tilde{c}_d c_0$, as the case of electro-osmotic instability (EOI) without geometrical confinement~\cite{1}. For moderate aspect ratios, $\tilde{H}^{-1}<10$, the dimensionless distance of the vortex center from the membrane, $\tilde{l}_c = l/L$, is nearly independent of voltage in the OLC regime (inset of Fig. 3b) and approximately equal to the dimensionless concentration, $\tilde{c}_d \approx \tilde{l}_c$.

Using this observation, we can derive a 1D model for the EOF regime~\cite{epaps} (inset of Fig. 3c) which yields an approximate OLC current-voltage relation,
\begin{equation}
\tilde {j} \approx
1-\tilde{c}_d^{2} +2\tilde{c}_d\ln \tilde{c}_d + \tilde{c}_d\tilde{V}
\ \ \mbox{ (EOF) }
\label{7}
\end{equation}
which compares well with our  2D simulations (Fig. 3c) for $\tilde{H}>0.1$ \cite{epaps}.  The linear dependence of (\ref{7}) again implies a constant shunt resistance (Fig. 2c) for the EOF mechanism, $\tilde{j}_s= \tilde{c}_d \tilde{V}$, as in (\ref{4}) for the SC mechanism, but with a different over-limiting conductance, $\tilde{c}_d$ rather than $-\tilde{\rho}_s$.

To understand the over-limiting conductance, we perform a scaling analysis of the 2D model~\cite{epaps}.  Balancing axial convection over the length scale $l_c$ with transverse diffusion over the scale $H$ in (\ref{5}), the vortex position scales as, $l_c\sim {u_{\textrm{EOF}}H^2}/D$. The characteristic EOF velocity along the walls scales as, $u_{EOF}\sim \tilde{\sigma}_s\tilde{j} u_{ref}/\tilde{c}^{3/2}\tilde{\lambda}_0$, where we use (\ref{6}) and define the reference (electroviscous) velocity scale, $u_{ref}=\varepsilon(kT/e)^2/\eta L$. The numerical observation, $\tilde{c}_d\approx\tilde{l}_c$, then implies the following scaling relation near limiting current ($\tilde{j}\approx 1$):
\begin{equation}
\tilde{c}_d\approx \tilde{l}_c \approx 0.22\, \tilde{\sigma}_s^{2/5}\tilde{H}^{4/5}{\tilde{\lambda}_0}^{-2/5}, \label{8}
\end{equation}
which shows how the over-limiting conductance due to EOF varies with surface charge, channel thickness, and reservoir concentration (via $\lambda_0$). Our 2D simulations (Fig. 4, inset) roughly confirm the scaling of $\tilde{c}_d \approx\tilde{l}_c$ with $\tilde{H}$ and provide the prefactor in (\ref{8}). For very large currents ($\tilde{j}\gg 1$), the conductance is predicted to increase~\cite{epaps} (Fig. 3c), but a detailed analysis also considering high aspect ratios ($\tilde{H}^{-1}\gg 10$) is left for future work.

\begin{figure}
\includegraphics[width=0.93\columnwidth,keepaspectratio=true]{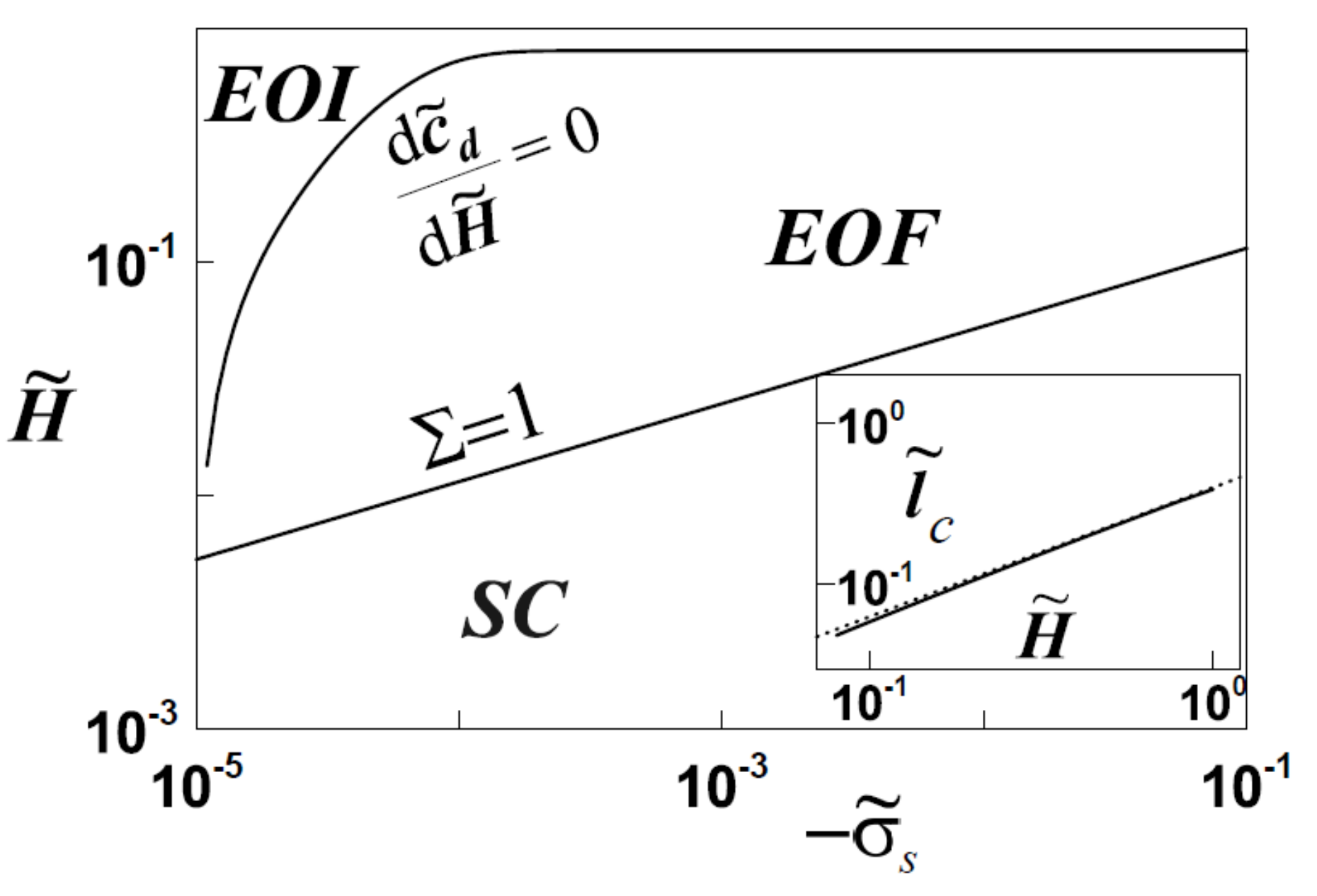}
\caption{ The regions of dominance of the SC, EOF, and EOI mechanisms for varying channel aspect ratio, $\tilde{H}^{-1}$ and surface charge $\tilde{\sigma}_s$ for $\tilde{\lambda}_0=10^{-5}$. The critical value of $\tilde{H}$ is given by $\Sigma=1$ in (\ref{9})  for the SC/EOF transition and by $\frac{d\tilde{c}_d}{d\tilde{H}}=0$ in 2D simulations for the EOF/EOI transition.
Inset: Simulated vortex center location $\tilde{l}_c$ versus $\tilde{H}$ for $\tilde{\lambda}_0=0.01$, $\tilde{\sigma}_s=-0.1$ (solid) compared to the scaling law, $\tilde{l}_c=0.38\tilde{H}^{4/5}$ in (\ref{8}).}
\end{figure}

{\it Unified scaling analysis. ---}
The relative importance of the SC and EOF mechanisms for OLC is controlled by the dimensionless ratio of their conductances in the weakly over-limiting regime,  $\Sigma=|{\tilde{\rho}_s}|/{\tilde {c}_d}$. Using (\ref{8}) and restoring dimensional variables, we obtain the scaling,
\begin{equation}
\Sigma\approx 8\tilde{\lambda}_0^{2/5}|\tilde{\sigma}_s|^{3/5}\tilde{H}^{-9/5}\sim H^{-9/5}L^{4/5}|\sigma_s^{3/5}|c_{0}^{-4/5}. \label{9}
\end{equation}
where $\Sigma=1$ marks the transition between mechanisms (Fig. 4).
Equation (\ref{9}) shows that SC dominates for thin channels, low ion concentration and/or large surface charge ($\Sigma \gg 1$), while EOF dominates above a critical channel thickness and/or salt concentration ($\Sigma \ll 1$).

As the channel thickness is increased further, surface effects eventually become negligible, and OLC results from bulk EOI at the membrane, as in unconfined systems~\cite{1,2}. The EOF vortex intensity, scaling as $\tilde{l_c}/\tilde{H}\sim H^{-1/5}$ from (\ref{8}), decreases with $\tilde H$, and thus, above a critical channel thickness, a significant portion of the depleted solution near the membrane is unaffected by EOF and instead undergoes EOI to maintain OLC.  EOI also leads to a constant over-limiting conductance (beyond a threshold voltage)~\cite{1,2}, so the EOF/EOI transition can be defined via $\frac{d\tilde{c}_d}{d\tilde{H}}=0$ in our 2D simulations, above which the EOF supply of solute to the membrane fades and conditions develop for EOI (Figs. 4 and S4 \cite{epaps}). In water, for $L=1$mm,  $\zeta_0=50$mV and $c_0=1$mM, the predicted mechanism for OLC transitions from SC to EOF at $H=8\mu$m (consistent with observed vortex supression~\cite{kim2007}) and from EOF to EOI at $H=0.4$mm, so that different microfluidic experiments can be dominated by SC~\cite{3}, EOF~\cite{han}, or EOI~\cite{1}.

\emph{Conclusion.} ---  In a microchannel, charged side walls can sustain OLC to an ion-selective end. With increasing channel thickness, the dominant mechanism switches from SC to EOF to EOI.  These nonlinear phenomena may find applications in water desalination, biological separations, or electrochemical energy storage.

This work was supported by a seed grant from the MIT Energy Initiative and by the Israel Science Foundation (Grant No. 65/07).

\clearpage

\centerline{\large{ \bf Supplementary Information }}

\vskip 12pt

\section{Effect of streaming current on surface conduction}
Equations (1) and (2) in the main text present a model for the area-averaged concentration in a channel. In this model axial transport of ions is considered via diffusion and electromigration including migration of excess ions shielding the wall charge, but the effects of flow is ignored (streaming current). We here present an estimate of this effect in the simplest form and show that consistent with our assumption, this effect is negligible for sufficiently thin dead end channels.

It should be noted that the relative importance of streaming current in surface conduction in the context of thin double layer is well established in the literature. For example, Deryagin and Dukhin$^{19}$ used the Gouy-Chapman EDL model and derived analytical expressions for streaming current carried by electroosmotic flow. Following a similar analysis one could show that the ratio of the streaming current to current carried by electro migration of excess counterions shielding the wall charge is:
\begin{equation}
 \frac{j_\text{flow}}{j_\text{cond.}} =2\text{Pe}\tanh\left(-\frac{ze\zeta}{4kT}\right),
\label{eq:thinEDL}
\end{equation}
where $\text{Pe}=\varepsilon(KT/ez)^2/\eta D$ is the material Peclet number and property of the solution. This analysis however, is not applicable to the case of channels considered in our studies; when channels themselves are thin, the thin EDL assumption will be violated due to overlap EDL effects. In addition, the pressure driven flow, which would cancel the net electroosmotic flow for dead-end channels, will contribute to current.

To remedy these issues we carried out a separate analysis specific to the case of dead-end thin channels. One can consider a channel with high aspect ratio ($L\gg H$); therefore the competing flow components, i.e. pressure driven and electroosmotic flow, can be assumed to be locally parallel. Furthermore, the EDLs (either overlapped or non-overlapped) can be assumed to be in local equilibrium$^{12}$. Therefore, the charge distribution can be estimated from solution of the Poisson-Boltzmann equation with top and bottom wall condition as known surface charge, $\sigma_s$. These solutions determine the distribution of the potential and charge in the wall-normal direction. Having such solutions one can determine the velocity profile associated with the electro-osmotic flow:
\begin{equation}
 u = -\frac{\varepsilon E}{\eta}\left(\phi(y)-\phi_\text{wall}\right),
\end{equation}
where $E$ is the axial electric field, and $\phi(y)-\phi_\text{wall}$ is the local potential difference in the wall-normal direction. The pressure driven flow is assumed to have a parabolic profile with a vanishing net flow. In such a scenario, near the walls the electro-osmotic flow dominates the pressure driven flow and advects the charged double layers forward; in the middle section, the pressure driven flow wins, but the available charge is much less than that near the wall. Therefore, the flow would contribute a net current to the system. This net current is proportional to the flow, which itself is proportional to the applied field, and thus has an Ohmic effect.

\begin{figure}
\includegraphics[width=2.5in,keepaspectratio=true]{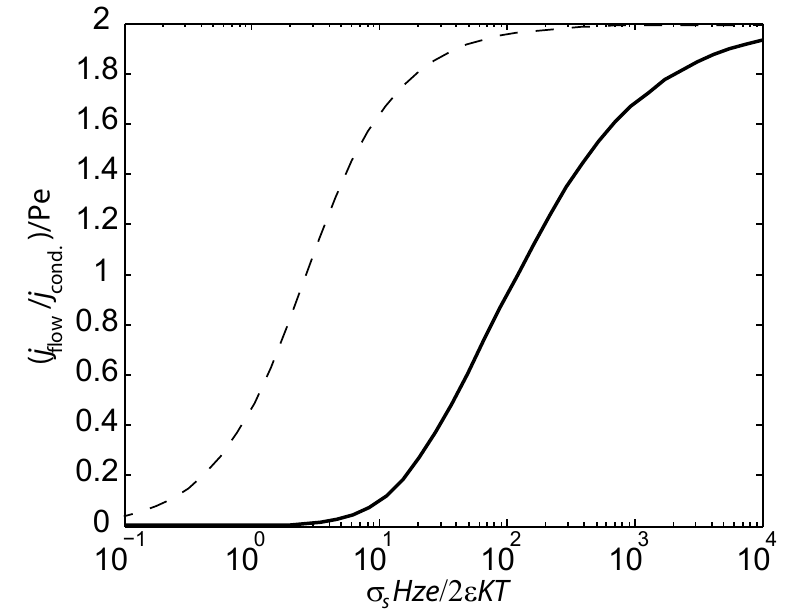}
\caption{Streaming current versus nondimensional wall charge. Solid line represent the result of our analysis for thin channels including effect of the pressure driven flow. Dashed line represents the thin-EDL approximation based on Eq.~\ref{eq:thinEDL}, for $\lambda/H=0.5$. Current is normalized by the current associated with electromigration of the surface shielding charge, $ez\nu|\sigma| E$ (per channel width). The plotted result is valid for all $\lambda_D/H>0.5$ within plotting accuracy.}
\label{fig1}
\end{figure}

This problem can be solved in the general form. The two dimensionless parameters associated with the Poisson-Boltzmann system are: $\lambda_D/H$, Debye length over channel thickness, and $\tilde{\rho}_s$, the dimensionless wall charge. The current contribution of the flow, becomes more important in the limit of highly depleted channel, i.e. $\lambda_D/H\rightarrow\infty$, and $\tilde{\rho}_s\rightarrow\infty$. In this limit, the meaningful controlling parameter is
\begin{equation}
\sigma^*=\frac{\sigma H ez}{2 \varepsilon KT}= \frac{\tilde{\rho}_sH^2}{4\lambda_D^2},
\end{equation}
which is finite and independent of the bulk concentration. Figure~\ref{fig1} shows the net streaming current as a function of $\sigma^*$ from our analysis and contrasts it to the prediction of thin EDL model (Eq.~\ref{eq:thinEDL}), which is clearly shown to overestimate the effect of streaming current.

For a one micron thin channel with surface charge $\sigma$=1mC, $\sigma^*$ will be about 30; assuming an aqueous solution at room temperature (Pe$\sim$0.5) the current contribution by the flow is estimated to be about 15\% of current by the electromigration of the surface charge. In this case ignoring the streaming current is marginally valid, and the approximation significantly improves as one uses thinner channels, however, for larger channels (or larger surface charge) the streaming current should be considered in the model (in simplest form, as a prefactor enhancing the surface conduction effect).

\section{EOF mechanism: scaling analysis}

Here we derive in more detail the scaling relation Eq.(9) for the over-limiting conductance due to electro-osmotic flow on the side-walls of the microchannel, which leads to a pressure-driven backflow from the impermeable membrane end of the microchannel.  The resulting vortex pair drives the fast convection of salty solution to the membrane along the walls and the return of depleted solution back into the microchannel (Fig. 2c). It is tempting to analyze convection-diffusion in this flow by area-averaging using the theory of Taylor-Aris dispersion, but we shall see that convective mixing in the transverse direction is not fully developed in the vortex region and leads to different scalings with system parameters.

Electro-osmotic flow near the channel wall scales as
\begin{equation}
 u \sim \frac{\varepsilon \zeta \phi_x }{\eta}
\end{equation}
where the zeta potential, $\zeta$, is related to the surface charge density using the thin diffuse-layer capacitance ($C = \varepsilon/\lambda_D$), and the tangential potential gradient, $\phi_x$,  is related to the current density, $j$:
\begin{equation}
\zeta \sim  \frac{\sigma_s \lambda_D}{\varepsilon}, \ \phi_x \sim \frac{ j }{ \kappa_b }
\end{equation}
where $\lambda_D(c)$ is the Debye length and $\kappa_b(c)$ is the bulk conductivity, each depending on the local bulk salt concentration $c$.

Combining these equations, we obtain the velocity scaling
\begin{equation}
u \sim \frac{\sigma_s \lambda_D j } {\kappa_b \eta }
\end{equation}

Near the dead end, the flow recirculates and produces a vortex which shears the fluid over a transverse scale $H$ (the channel thickness) and longitudinal scale $\l_c$ (the vortex center). The concentration is strongly depleted by the selective surface (e.g. membrane) near the end of the channel, where the electric field driving the flow is at maximum. Near this end the flow turns and, upon moving away from the end, eventually develops into a plane parallel slip/pressure driven flow for which the vortex center location is similar to an entrance length. This result is reminiscent of the convection-diffusion problem for the evaluation of the entrance length of a cold fluid entering a hot parallel-plate pipe. In our terms, the length of the  diffusion scale is the channel thickness $H$, while the longitudinal convection scale is the vortex center location (the entrance length) $\l_c$. The convection-diffusion equation, $\vec{u}\cdot\nabla c = D \nabla^2 c$, yields the scaling
\begin{equation}
\frac{u}{\l_c} \sim \frac{D}{H^2}
\end{equation}
which implies
\begin{equation}
\l_c \sim \frac{u H^2}{D} \sim \frac{\sigma_s \lambda_D j H^2 } {\kappa_b \eta  D }
\end{equation}

Next, we introduce dimensionless quantities, denoted by tilde accents. Let us employ the channel length $L$ for the length scale, $kT/ze$ for the potential, and thus the electro-viscous scale,
\begin{equation}
u_{ref} = \frac{ \varepsilon (kT/ze)^2}{\eta L}
\end{equation}
for the velocity scale. Let us also scale concentration to the bulk reservoir concentration $c_0$ at the open end of the channel. Assume an $z:z$ electrolyte with equal diffusivities and an ideally cation selective surface.  This yields the following expression for conductivity and Debye length,
\begin{equation}
\kappa_b = \frac{\varepsilon D}{\lambda_D^2}, \ \lambda_D^2 = \frac{ \varepsilon k T }{2(ze)^2 c}
\end{equation}
The dimensionless values for current, height and surface charge, are defined as
\begin{equation}
\tilde{j} = \frac{j L}{2 z e D c_0 }, \ \tilde{H} = \frac{H}{L}, \ \tilde{\sigma}_s = \frac{\sigma_s }{ze c_0 L}
\end{equation}
This nondimensionalization yields two crucial parameters the ``material P\'eclet number", \mbox{Pe}, ($\approx 1/2$ for aqueous solutions at room temperature) and dimensionless Debye length, $\tilde{\lambda}_0$, at the entrance to the channel, defined as
\begin{equation}
\mbox{Pe} = \frac{ u_{ref} L}{D} = \frac{\varepsilon (kT/ze)^2 }{\eta D}, \ \tilde{\lambda}_0^2 = \frac{ \varepsilon k T }{2(ze)^2 c_0 L^2}
\end{equation}

With these definitions, the scaling of the vortex center takes the dimensionless form
\begin{equation}
\tilde{\l}_c \sim \frac{\mbox{Pe} \, \tilde{H}^2 \tilde{\sigma}_s\tilde{j}}{\tilde{\lambda}_0 \tilde{c}^{3/2} } \label{eq:scale}
\end{equation}

The only unknown in the scaling (\ref{eq:scale}) is the dimensionless concentration in the mixed depleted region. Using the numerically determined relationship
\begin{equation}
\tilde{c}_d \sim \tilde{\l}_c   \label{eq:c}
\end{equation}
which is valid in the weakly overlimiting regime (Section 2 below), we obtain
\begin{equation}
\tilde{\l}_c \sim \left( \frac{ \mbox{Pe} \, \tilde{H}^2 \tilde{\sigma}_s }{\tilde{\lambda}_0} \right)^{2/5}   \label{eq:l}
\end{equation}
where the coefficient of proportionality is approximately 0.3 from simulations. As this equation is applicable only in the weakly overlimiting regime ($\tilde{j}\sim 1$), the current density does not affect the scaling of the overlimiting conductance. If we were to keep the current from our derivation and still use (\ref{eq:c}), then we would obtain
\begin{equation}
\tilde{c}_d \sim \left( \frac{ \mbox{Pe} \, \tilde{H}^2 \tilde{\sigma}_s \tilde{j} }{\tilde{\lambda}_0}  \right)^{2/5}   \label{eq:clarge}
\end{equation}
for very large currents, $\tilde{j} \gg 1$, in place of (\ref{eq:l}).

To compare the EOF vortex mechanism for over-limiting current (dominant for wide channels) to the mechanism of surface conduction (dominant in narrow channels) in the weakly overlimiting regime, we consider the dimensionless parameter $\Sigma$,
\begin{equation}
\Sigma = \frac{\tilde{\rho}_s }{\tilde{c}_d} = \frac{\tilde{\sigma}_s}{\tilde{c}_d \tilde{H} }
\end{equation}
which is the ratio of the two different possible depth-averaged conductivities in the depleted region, where
\begin{equation}
\tilde{\rho}_s = \frac{\sigma_s}{ze c_0 H} = \frac{\tilde{\sigma}_s}{ \tilde{H} }
\end{equation}
is the scaling of the limiting conductivity due to surface conduction. Using Eqs.~(\ref{eq:c})-(\ref{eq:l}) and the simulation results, we obtain
\begin{equation}
\Sigma \approx 6   \left( \frac{ \tilde{\lambda}_0^2 \tilde{\sigma}_s^3 }{\mbox{Pe}^2 \tilde{H}^9 } \right)^{1/5}
\end{equation}
which depends on the channel dimensions, surface charge density, and reservoir salt concentration as
\begin{equation}
\Sigma  \sim H^{-9/5}  L^{4/5} \sigma_s^{3/5} c_0^{-4/5}.
\end{equation}
For sufficiently narrow and long channels, thick double layers, large surface charge density, and/or small reservoir concentration, $\Sigma \gg 1$, surface conduction is the dominant mechanism for over-limiting current. In the opposite limit, $\Sigma \ll 1$, the dominant mechanism is convective mixing by confined electro-osmotic  flow.

\begin{figure}
\includegraphics[width=\columnwidth,keepaspectratio=true]{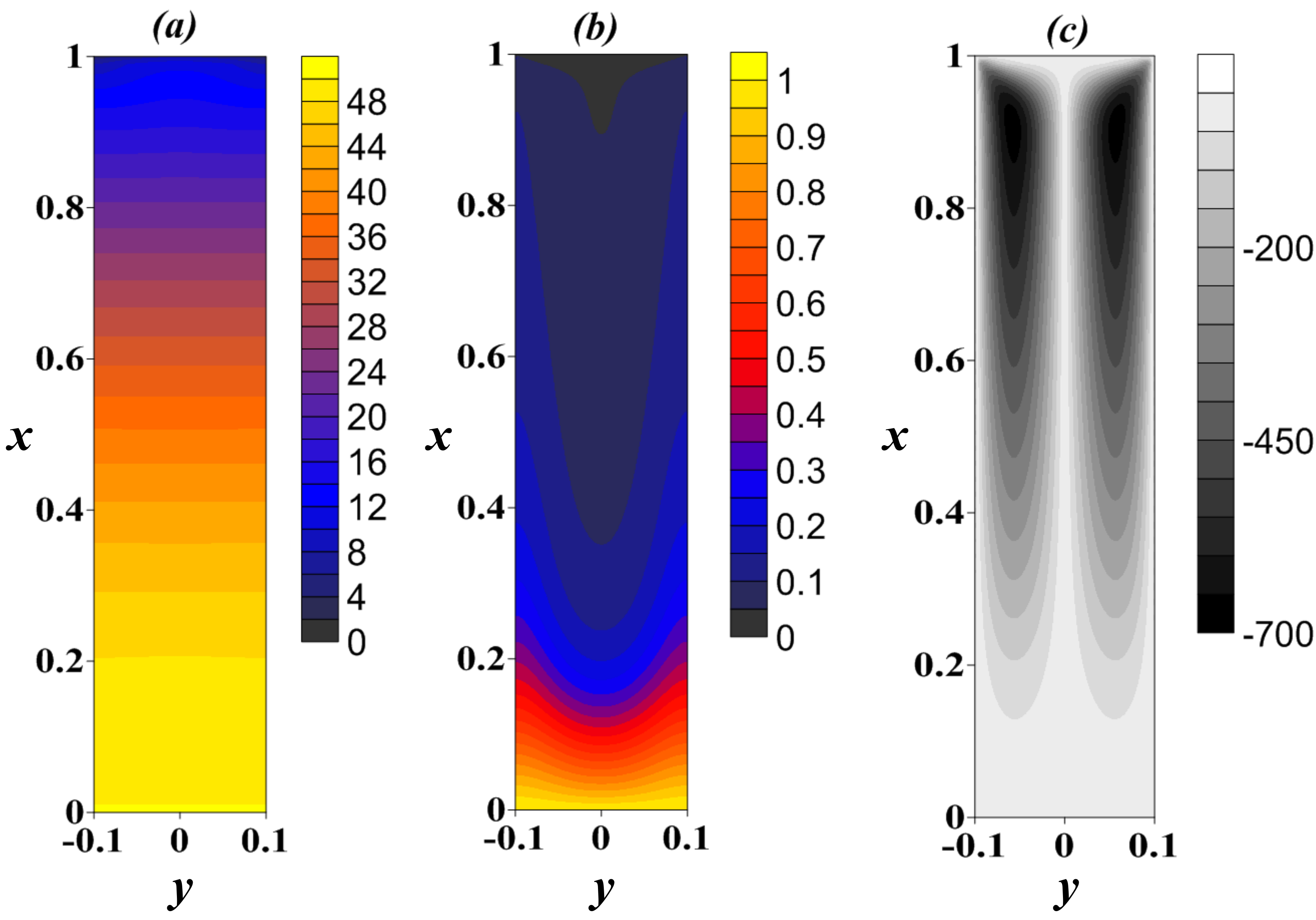}
\caption{(a) Electric potential, $\tilde{\varphi}$, level lines;  (b) concentration, $\tilde{c}$, level lines and (c) EOF streamlines computed for $\tilde {N}=10$, $\tilde{\lambda}_0=0.01$, $\tilde{\sigma}_s=0.01$, $\tilde{V}=50$  and $\tilde{H}=0.2$.}
\end{figure}

\begin{figure*}
\includegraphics[width=\linewidth,keepaspectratio=true]{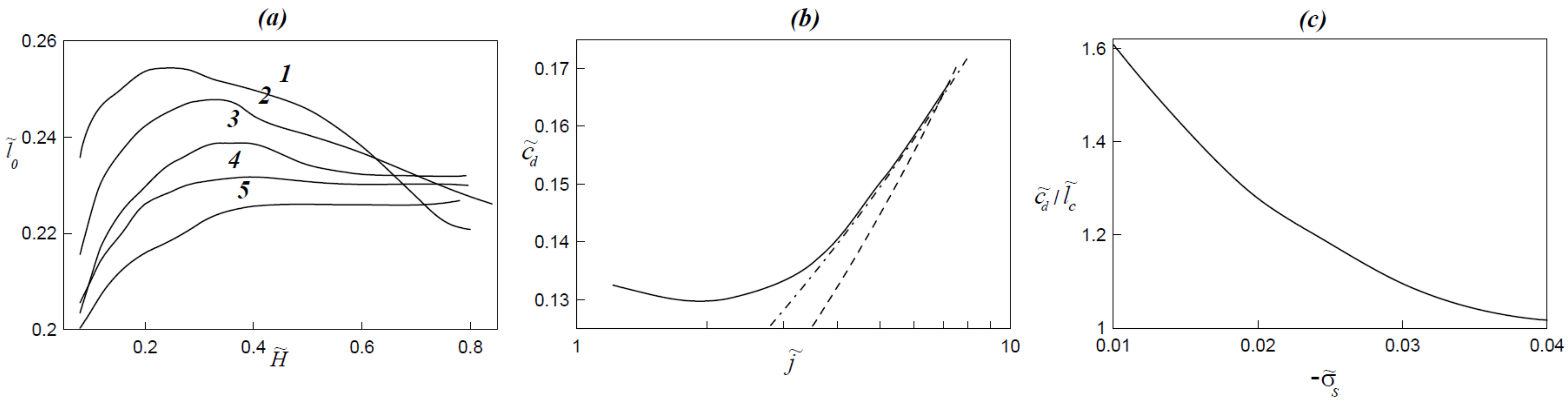}
\caption{Simulation of EOF mechanism for $\tilde{\lambda}_0=0.01$: (a) The coefficient of the proportionality $\tilde{l}_0$ versus channel depth $\tilde{H}$ for $\tilde{j}=1.5$ and the following values of the surface charge density: (1) -- $\tilde{\sigma}_s=0.02$, (2) -- $\tilde{\sigma}_s=0.025$, (3) -- $\tilde{\sigma}_s=0.03$, (4) -- $\tilde{\sigma}_s=0.035$, (5) -- $\tilde{\sigma}_s=0.04$; (b) Concentration of the depleted bulk $\tilde{c}_d$ versus the electric current density $\tilde{j}$ for $\tilde{H}=0.2$ and $\tilde{\sigma}_s=0.04$, compared with $\tilde{j}^{1/3}$ and $\tilde{j}^{2/5}$ asymptotic power laws (dashed lines); (c) the ratio $\tilde{c}_d/\tilde{l}_c$ versus surface charge density $\tilde{\sigma}_s$ computed for $\tilde{j}=1.5$ and  $\tilde{H}=0.2$.}
\end{figure*}

\section{ 2D Simulations }

Here we study in detail the EOF mechanism of the OLC in the dead-end microchannel via 2D numerical simulations in the limit of thin double layers. The relevant set of steady-state dimensionless Nernst-Planck-Stokes equations for the electroneutral bulk, neglecting SC, reads
\begin{align}
& \textrm{Pe}\mathbf{\tilde{u}}\cdot\mathbf{\nabla} \tilde{c}=\mathbf{\nabla}^2 \tilde{c},\ \mathbf{\nabla}(\tilde{c}\mathbf{\nabla}\tilde{\varphi})=0, \ \mathbf{\nabla}^2\mathbf{\tilde{u}}-\mathbf{\nabla}p=0;\label{1}\\
& y=0,\ \tilde{H}:\ \frac{\partial \tilde{c}}{\partial\tilde{y}}=\frac{\partial \tilde{\varphi}}{\partial\tilde{y}}=0, \mathbf{\tilde{u}}\cdot \mathbf{n}=0,\ \mathbf{\tilde{u}}\cdot \mathbf{\tau}=\tilde{\zeta}\frac{\partial\tilde{\varphi}}{\partial \tilde {x}}\label{2},\\
& x=1:\ \frac{\partial \tilde{c}}{\partial\tilde{x}}-\tilde{c}\frac{\partial \tilde{\varphi}}{\partial\tilde{x}}=0, \mathbf{\tilde{u}}=0,\ \ln \tilde{c}+\tilde{\varphi}=\ln \tilde{N};\label{3}\\
& x=0:\ \tilde{c}=1,\ \tilde{\varphi}=\tilde{V},\ \mathbf{\tilde{u}}\cdot\mathbf{\tau}=0,\ \frac{\partial}{\partial \tilde x}(\mathbf{\tilde{u}}\cdot\mathbf{n})=0.\label{4}\
\end{align}
Here $\tilde\zeta$ in the Helmholtz--Smoluchowsky slip condition (\ref{2}) is the dimensionless $\zeta$-potential, which we assume is related to the surface charge of the nonconducting side wall through the linear (low-voltage) Debye--Huckel relation,
\begin{equation}
\tilde{\zeta}=\frac{\tilde{\sigma_s}}{\tilde{\lambda}_0\sqrt{\tilde{c}}};\label{5}
\end{equation}
neglecting any effects of the compact Stern layer.
Conditions (\ref{3}) at the perfect cation-selective membrane/solution interface are those of vanishing anions' flux  and continuity of the electrochemical potential of cations across the interface, along with the non-slip condition; $\tilde{N}$ is the dimensionless fixed charge density in the membrane and $\tilde{V}$ is once more the potential drop across the diffusion layer.

Below we present the results of numerical solution of the problem (\ref{1}--\ref{4}).
Thus, in Fig.2 we present the concentration and electric potential level lines and EOF streamlines obtained for highly OLC regime.
In Fig.3 we show the results of numerical verification of scaling relations (Eq.(9), main text) derived in the main part of this Letter. Thus, in Fig.3a we present the numerically computed coefficient of proportionality $\tilde{l}_0$
\begin{equation}
\tilde{l}_0=\frac{\tilde{l}_c}{{\tilde{\sigma}_s}^{2/5}\tilde{H}^{4/5}\tilde{\lambda}_0^{-2/5}}
\end{equation}
versus channel depth $\tilde{H}$ for a sequence of values of surface charge density $\tilde{\sigma}_s$. In Fig.3b we show the weak dependence of the depleted bulk concentration $\tilde{c}_d$ on the electric current density in the weakly overlimiting current regime, which transitions to  power-law dependence in the strongly over-limiting regime. Although the simulations may be consistent with an asymptotic power law, $\tilde{c}_d\sim\tilde{j}^{2/5}$, as predicted by our simple scaling analysis, Eq.~(\ref{eq:clarge}), a somewhat weaker dependence, $\tilde{c}_d\sim\tilde{j}^{1/3}$, provides a better fit over the range of moderately large currents in the simulations (up to $\tilde{j}=6$).
In Fig.3c we show that the increasing rate of the EOF due to an increase of surface charge density yields convergence of the dimensionless concentration of the depleted bulk $\tilde{c}_d$ to the dimensionless location of the vortex center $\tilde{l}_c$ for the weakly overlimiting current regime.

Finally, to illustrate the transition to EOI mediated OLC with the decrease of the channel aspect ratio in Fig. 4a we present  plot of the dimensionless concentration of the depleted bulk $\tilde{c}_d$ versus $\tilde{H}$ which shows a sharp decrease of $\tilde {c}_d$ for $\tilde{H}$ above some threshold due to the increasing insensitivity of the middle part of the channel to the wall-induced EOF. In Fig. 4b,c we illustrate this by presenting the EOF streamline maps for $\tilde H=0.2$ below and $\tilde H=1.6$, above this threshold.

\begin{figure}
\includegraphics[width=\columnwidth,keepaspectratio=true]{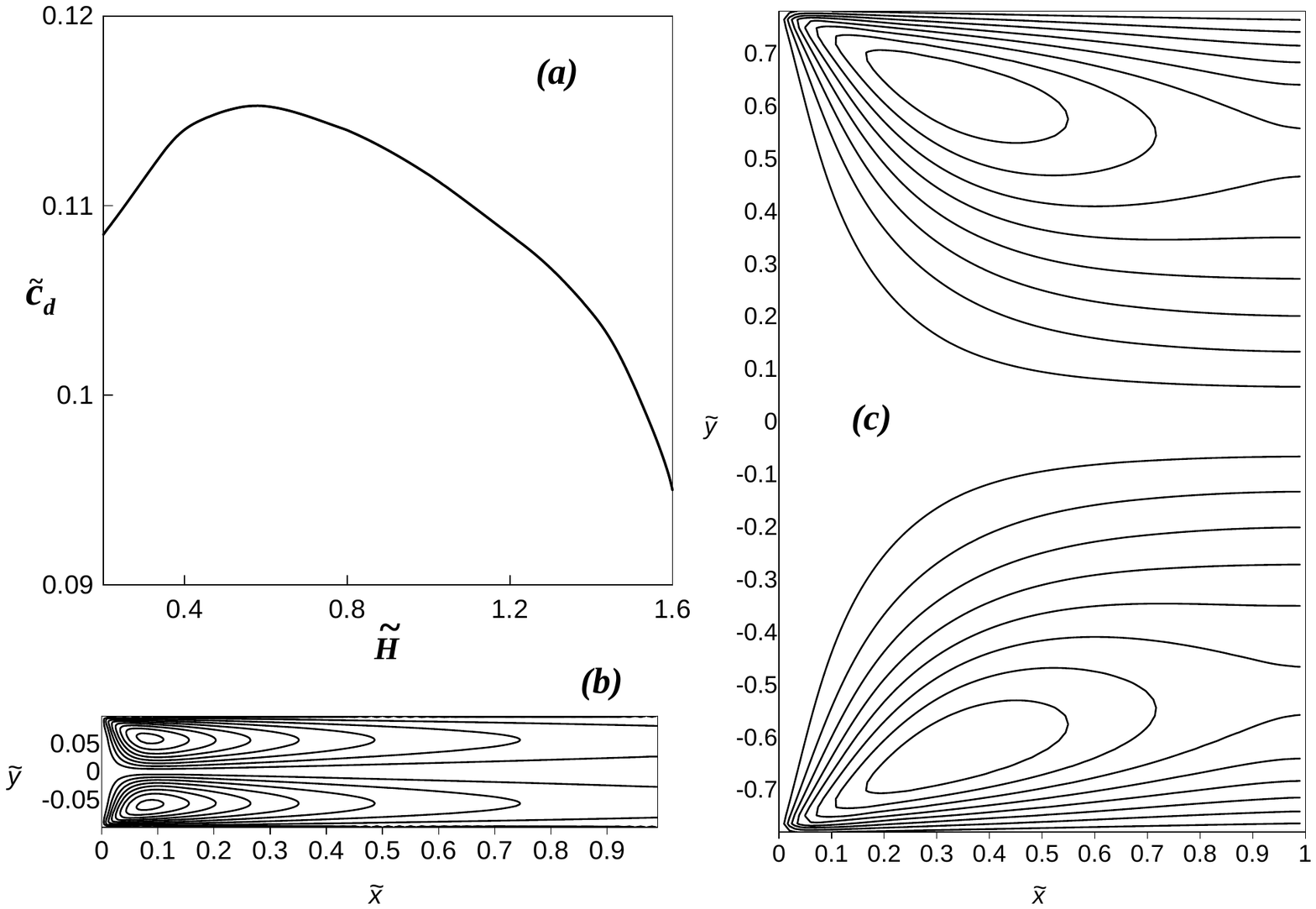}
\caption{(a) The dimensionless concentration of the depleted bulk $\tilde{c}_d$ versus $\tilde{H}$ for $\tilde{\sigma}_s=10^{-3}$ and $\tilde V=10$;  (b) EOF streamlines for $\tilde{\sigma}_s=10^{-3}$, $\tilde V=10$ and $\tilde H=0.2$; (c) EOF streamlines for $\tilde{\sigma}_s=10^{-3}$, $\tilde V=10$ and $\tilde H=1.6$.}
\end{figure}

\section{ Approximate 1D model }

Motivated by our 2D simulations, we can develop a simple 1D phenomenological model for the EOF driven OLC in which $\tilde{l}_c$ is the sole fitting parameter. In this model the dimensionless salt concentration in the depleted region is fixed at $\tilde{c}_d=\tilde{l}_c$, whereas in the rest of the diffusion layer it varies linearly from $\tilde{l}_c$ to unity at the open end of the channel and from $\tilde{l}_c$ to $\tilde{l}_c^2$ at the dead end of the channel:
\begin{align}
&\tilde{c}=\tilde{l}_c,\ \frac{1-\tilde{l}_c}{\tilde{j}}<\tilde{x}<1-\frac{\tilde{l}_c-\tilde{l}_c^2}{\tilde{j}};\\
& \tilde{c}=1-\tilde{j}\tilde{x},\ 0<\tilde{x}<\frac{1-\tilde{l}_c}{\tilde{j}};\\
&\tilde{c}=\tilde{l}_c^2+\tilde{j}(1-\tilde{x}),\ 1-\frac{\tilde{l}_c-\tilde{l}_c^2}{\tilde{j}}<\tilde{x}<1;\\
&\tilde{c}\frac{d\tilde{\phi}}{d\tilde{x}}=-\tilde{j},\ 0<x<1, \tilde{\phi}(1)-\tilde{\phi}(0)=-\tilde{V}.\
\end{align}
The resulting current/voltage relation
\begin{equation}
\tilde{V}=\frac{\tilde{l}_c^2 - 1}{\tilde{l}_c} - 2 \ln \tilde{l}_c + \frac{\tilde{j}}{\tilde{l}_c},
\end{equation}
is plotted in Fig.3c (main text) assuming constant $\tilde{c}_d\approx \tilde{l}_c$. Comparison with the 2D simulations allows us to roughly confirm the scaling relation (\ref{eq:l}) and fit the prefactor.

We can also begin to understand the appearance of a weak nonlinearity in the 2D simulation results (Fig . 3c, main text), where the over-limiting conductance due to EOF grows slowly with voltage.
Taking into account the weak variation in $\tilde{c}_d$  for $\tilde{j}$ from the scaling analysis,  we would predict a weak nonlinearity, $\tilde{j} \sim \tilde{V}^{5/3}$ for very large currents, $\tilde{j} \gg 1$. Alternatively, assuming $\tilde{c}_d \sim \tilde{j}^{1/3}$ from Fig. 2b would suggest $\tilde{j} \sim \tilde{V}^{4/3}$, which may also be consistent with the simulations. A detailed analysis of the strongly over-limiting regime is left for future work.

\end{document}